\documentclass{PoS}

\usepackage{graphicx, fancybox}
\usepackage{amsmath, amssymb, mathrsfs, bm}
\usepackage[T1]{fontenc}
\usepackage[table]{xcolor}
\usepackage{ulem}
\usepackage{cite}

\newcommand{\red}[1]{\textcolor{red}{#1}}
\newcommand{\blue}[1]{\textcolor{blue}{#1}}

\title{
\begin{picture}(0,0)(0,0)%
 \put(135,90){\makebox(0,0)[l]{\textnormal{\normalsize 
 J-PARC-TH-0137, RIKEN-iTHEMS-Report-18, RIKEN-QHP-384}}}%
\end{picture}%
   Linear confinement and stress-energy tensor \\
   around static quark and anti-quark pair \\
   -- Lattice simulation with Yang-Mills gradient flow --}
   
\ShortTitle{Linear confinement and stress-energy tensor
around static quark and anti-quark pair}

\author{\speaker{Ryosuke Yanagihara},$^a$ Takumi Iritani,$^b$
Masakiyo Kitazawa,$^{a,c}$ Masayuki Asakawa,$^a$ Tetsuo Hatsuda$^{d,b}$
(FlowQCD collaboration)
\\
         \llap{$^a$} Department of Physics, Osaka University, Toyonaka,
	 Osaka 560-0043, Japan
	 \\
	 \llap{$^b$}RIKEN Nishina Center, RIKEN, Wako, 351-0198, Japan
	 \\
	 \llap{$^c$}J-PARC Branch, KEK Theory Center, Institute of Particle
	 and Nuclear Studies, KEK, 203-1, Shirakata, Tokai,
	 Ibaraki 319-1106, Japan
	 \\
	 \llap{$^d$} RIKEN Interdisciplinary Theoretical and Mathematical
	 Sciences Program (iTHEMS), RIKEN, Wako 351-0198, Japan
	 \\
         E-mail: \email{yanagihara@kern.phys.sci.osaka-u.ac.jp}}

	 \abstract{We study the spatial distribution of the stress tensor
	 around static quark-anti-quark pair in SU(3) lattice gauge theory.
	 In particular, we reveal the transverse structure
	 of the stress tensor distribution in detail
	 by taking the continuum limit.
	 The Yang-Mills gradient flow plays a crucial role
	 to make the stress tensor well-defined and derivable
	 from the numerical simulations on the lattice
         \cite{Yanagihara:2018qqg}.}

	 \FullConference{The 36th Annual International Symposium
	 on Lattice Field Theory - LATTICE2018\\
		22-28 July, 2018\\
		Michigan State University, East Lansing, Michigan, USA.}

\begin{document}

\section{Introduction}

The energy-momentum tensor (EMT), $\mathcal{T}_{\mu \nu}(x)$,
is a fundamental observable in physics.
Its spatial component is called the stress tensor,
$\sigma_{ij}=-\mathcal{T}_{ij}$, which represents 
the distortion of fields induced by external sources.
The Maxwell stress-tensor 
in electromagnetism is a well-known example:
$\sigma_{ij}^\mathrm{Maxwell}=-\mathcal{T}_{ij}^\mathrm{Maxwell}
=  -(F_{i\mu} F^{\mu}_{\ j} - \frac{1}{4} \delta_{ij} F_{\mu\nu}^2)$
~\cite{Landau}.
EMT in Yang-Mills (YM) theory 
plays a particularly important role because it represents the response
of the YM field against external sources in a gauge invariant manner.

In this work~\cite{Yanagihara:2018qqg}, we explore the stress
distribution around a static quark ($Q$)
and an anti-quark ($\bar{Q}$) separated by length $R$.
In YM theory, the field strength is believed to be squeezed
into a quasi-one-dimensional structure called the flux tube 
and gives rise to the linear confining potential~\cite{Bali:2000gf}.
In previous studies on the structure of the flux tube,
the action density and the color electric field have been
employed~\cite{Bali:1994de,Cea:2012qw,Cardoso:2013lla}.
In the present study we perform a first measurement of the 
stress tensor distribution around static $Q\bar{Q}$ pair.
We use the EMT operator on the lattice~\cite{Suzuki:2013gza} 
defined via the YM gradient flow~\cite{Luscher:2010iy},
which has been applied extensively to the study on
thermodynamics~\cite{Asakawa:2013laa,Kitazawa:2016dsl}.

\section{Energy-Momentum Tensor around Static Quark and Anti-Quark}

We first recapitulate 
the general feature of ${\cal T}_{\mu \nu}(x)$ in the Euclidean spacetime
with $\mu,\nu=1,2,3,4$.
The local energy density and  the stress tensor read respectively as 
$\varepsilon (x) = -  \mathcal{T}_{44}(x),\,
\sigma_{ij} (x) = - \mathcal{T}_{ij}(x)\ (i,j=1,2,3)$.
The force per unit area $\mathcal{F}_i$
is the momentum flow through a given
surface element with the normal vector $n_i$:  
$\mathcal{F}_i =  \sigma_{ij}n_j =  - {\cal T}_{ij} n_j$ \cite{Landau}.
The principal axes of stress tensor is obtained by diagonalized
${\cal T}_{ij}$ as 
${\cal T}_{ij}n_j^{(k)}=\lambda_k n_i^{(k)} \, (k=1,2,3)$, 
where $n_i^{(k)}$ are the principal axes and 
the strengths of the force per unit area
along $n_i^{(k)}$
are given by the absolute values of the
eigenvalues, $\lambda_k $.
The force acting on a test charge is
obtained by the surface integral ${F}_i =- \int_S {\cal T}_{ij} dS_j$, where $S$ is a surface surrounding the charge with the surface vector $S_j$ oriented  outward from $S$.

Next let us review how we obtain ${\cal T}_{\mu \nu}(x)$
non-perturbatively around static $Q\bar{Q}$ on the lattice.
First, 
a static $Q\bar{Q}$ system on the lattice is prepared
with the rectangular Wilson loop $W(R,T)$ 
where $Q\bar{Q}$ locate at
$\vec{R}_{\pm}=(0,0, \pm  R/2)$ and in the temporal interval $[-T/2, T/2]$.
Second, to define EMT in YM theory we use the YM gradient flow
\cite{Suzuki:2013gza,Asakawa:2013laa,Kitazawa:2016dsl}.
The starting point of this method is the YM gradient flow 
equation
$dA_\mu(t,x)/dt
= -g_0^2 \delta S_{\mathrm{YM}}(t)/\delta A_\mu (t,x)$,
where $t$ denotes the fictitious 5-th dimensional coordinate~\cite{Luscher:2010iy}, and 
$S_{\mathrm{YM}}(t)$
is composed of $A_{\mu}(t,x)$,
whose initial condition at $t=0$ is
the ordinary gauge field $A_\mu(x)$ in the four dimensional
Euclidean space.
The gradient flow for positive $t$ leads to cooling of the gauge field
within the radius $ \sqrt{2t}$.
From the flowed field, 
the renormalized EMT operator is defined as ~\cite{Suzuki:2013gza}
\begin{align}
 {\cal T}^{\rm R}_{\mu\nu}(x)=\lim_{t\rightarrow0}{\cal T}_{\mu\nu}(t,x),
 \quad
 {\cal T}_{\mu\nu}(t,x)=\frac{U_{\mu\nu}(t,x)}{\alpha_U(t)}
 +\frac{\delta_{\mu\nu}}{4\alpha_E(t)}  [E(t,x)- \langle E(t,x)
   \rangle_0 ].
 \label{eq:T}
\end{align}
Here $E(t,x)=(1/4)G_{\mu\nu}^a(t,x)G_{\mu\nu}^a(t,x)$ and
$U_{\mu\nu}(t,x)=G_{\mu\rho}^a(t,x)G_{\nu\rho}^a(t,x)-\delta_{\mu\nu}E(t,x)$
with the field strength $G_{\mu\nu}^a(t,x)$ composed of the flowed gauge field $A_\mu (t,x)$.
The vacuum expectation value $\langle {\cal T}_{\mu\nu}(t,x) \rangle_0$ is normalized to be zero due to the subtraction of $\langle E(t,x) \rangle_0$.
We use the next-to-leading-order coefficients for $\alpha_U(t)$ and
$\alpha_E(t)$~\cite{Suzuki:2013gza}.
The validity and usefulness of this EMT operator
have been confirmed via the study on thermodynamic quantities in SU(3) YM 
theory~\cite{Asakawa:2013laa,Kitazawa:2016dsl}.

The expectation value of ${\cal T}_{\mu\nu}(t,x)$ around the $Q\bar{Q}$ 
is obtained by \cite{Luscher:1981}
\begin{align}
 \langle {\cal T}_{\mu\nu}(t,x)\rangle_{Q\bar{Q}} =
 \lim_{T\to\infty} \frac{\langle  {\cal T}_{\mu\nu}(t,x) W(R,T)\rangle_0}
 {\langle W(R,T)\rangle_0} ,
 \label{eq:<T>_W} 
\end{align}
where $T \rightarrow \infty$ is to pick up the ground state of
$Q\bar{Q}$.
The measurements of ${\cal T}_{\mu\nu}(t,x)$ for different values of $t$
are made at the mid temporal plane $x_{\mu} =(\vec{x}, x_4=0)$,
while $W(R,T)$ is defined at $t=0$.

Finally, we obtain the renormalized EMT distribution around
${Q\bar{Q}} $ from the lattice data by taking the double limit, 
$\langle {\cal T}^{\rm R}_{\mu\nu}(x)\rangle_{Q\bar{Q}}
  = \lim_{t \rightarrow 0}  \lim_{a \rightarrow 0}
  \langle {\cal T}_{\mu\nu}(t,x)\rangle_{Q\bar{Q}}^{\rm lat}$
  ~\cite{Asakawa:2013laa,Kitazawa:2016dsl}.
In lattice simulations we measure
$\langle {\cal T}_{\mu\nu}(t,x)\rangle_{Q\bar{Q}}^{\rm lat}$ at finite $t$ and $a$, and 
make an extrapolation to $(t,a)=(0,0)$ according to the formula~\cite{Kitazawa:2016dsl},
$\langle {\cal T}_{\mu\nu}(t,x)\rangle_{Q\bar{Q}}^{\rm lat}
\simeq  \langle {\cal T}^{\rm R}_{\mu\nu}(x)\rangle_{Q\bar{Q}} + b_{\mu
\nu}(t){a^2} + c_{\mu \nu} t$, 
where $b_{\mu \nu}(t)$ and $c_{\mu \nu} $ are contributions from
lattice discretization effects and the dimension six
operators, respectively.

\begin{table*}
 \centering
\begin{tabular}{|cccr|ccc|c|}
 \hline \hline
 $\beta$ & $a~[\mathrm{fm}]$ & $N_\mathrm{size}^4$ & $N_{\rm conf}$
 &    & $R/{a}$  &  & $T/{a}$     \\
 \hline
 6.304 & 0.058 & 48$^4$ & 140   & 8  & 12 & 16 & 8    \\
 6.465 & 0.046 & 48$^4$ & 440    & 10 & -- & 20 & 10 \\
 6.513 & 0.043 & 48$^4$ & 600    & -- & 16 & -- & 10 \\
 6.600 & 0.038 & 48$^4$ & 1,500  & 12 & 18 & 24 & 12 \\
 6.819 & 0.029 & 64$^4$ & 1,000 & 16 & 24 & 32 & 16   \\
 \hline

   & & \multicolumn{2}{c|}{$R ~[\mathrm{fm}]$}  &   0.46 & 0.69  & 0.92 &    \\
 \hline \hline
 \end{tabular}
 \caption{
   Simulation parameters on the lattice~\cite{Yanagihara:2018qqg}.
   Spatial size of the Wilson loop $R$ 
     is shown in the lattice and physical units.
   Temporal size of the Wilson loop is set to $T$
     unless otherwise stated.
 }
 \label{table:param}
\end{table*}

\section{Setup}
We have performed the numerical simulations 
in SU(3) YM theory on the four-dimensional Euclidean lattice 
with the Wilson gauge action
and the periodic boundary condition.
The setup for the simulation is summarized in Table~\ref{table:param}.
In the measurement of the Wilson loop $W(R,T)$,
we adopt the standard APE smearing for each spatial link 
~\cite{Albanese:1987ds}
to enhance the overlap between $W(R,T)$ and the $Q\bar{Q}$ ground state.
For the noise reduction, we also adopt the
standard multi-hit procedure by replacing every temporal links~\cite{Parisi:1983hm,Bali:1994de}.

\begin{figure}[t]
 \centering
 \includegraphics[width=0.99\textwidth,clip]{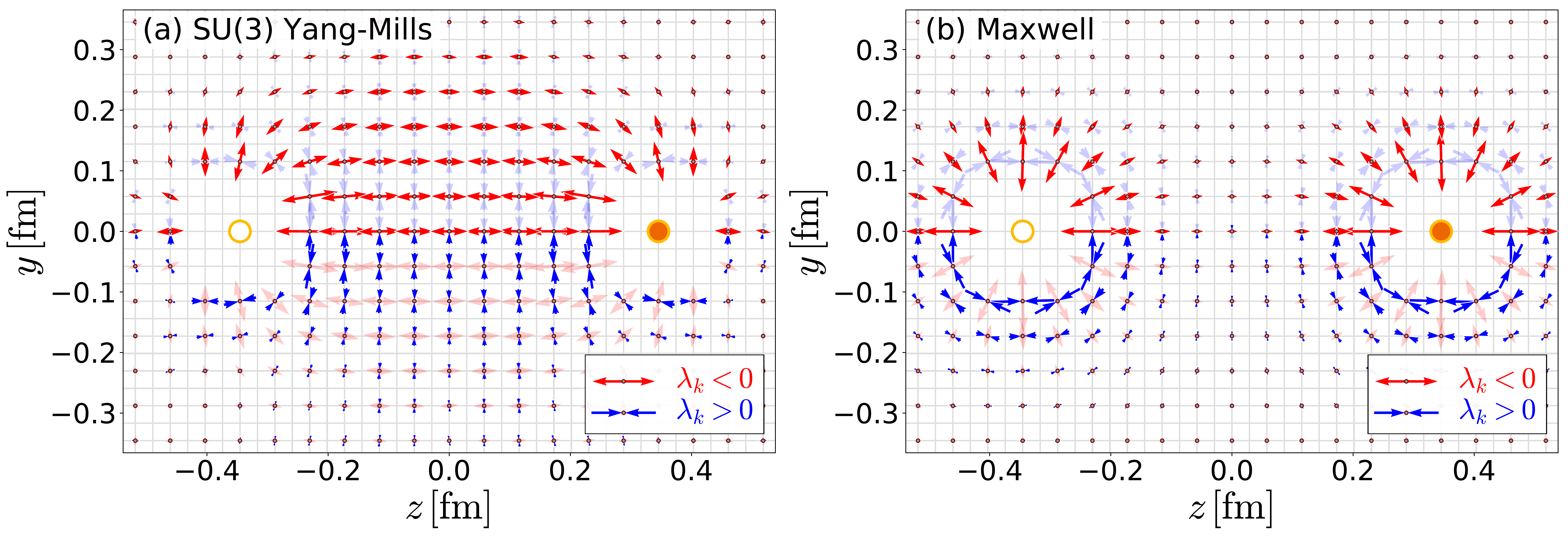}
 \caption{(a) Distribution of the principal axes of ${\cal T}_{ij}$ for a
 $Q\bar{Q}$ system separated by $R=0.69$ fm in SU(3) Yang-Mills theory 
 with $a=0.029$~fm and $t/a^2=2.0$~\cite{Yanagihara:2018qqg}.
 (b) Distribution of the principal axes of the ${\cal T}_{ij}$
 in classical electrodynamics between opposite charges.
 In both figures, the red (blue) arrows in the upper (lower) half plane are 
 highlighted. }
 \label{fig:stress-distribution}
\end{figure}

\begin{figure*}[t]
 \centering
 \includegraphics[width=0.99\textwidth,clip]{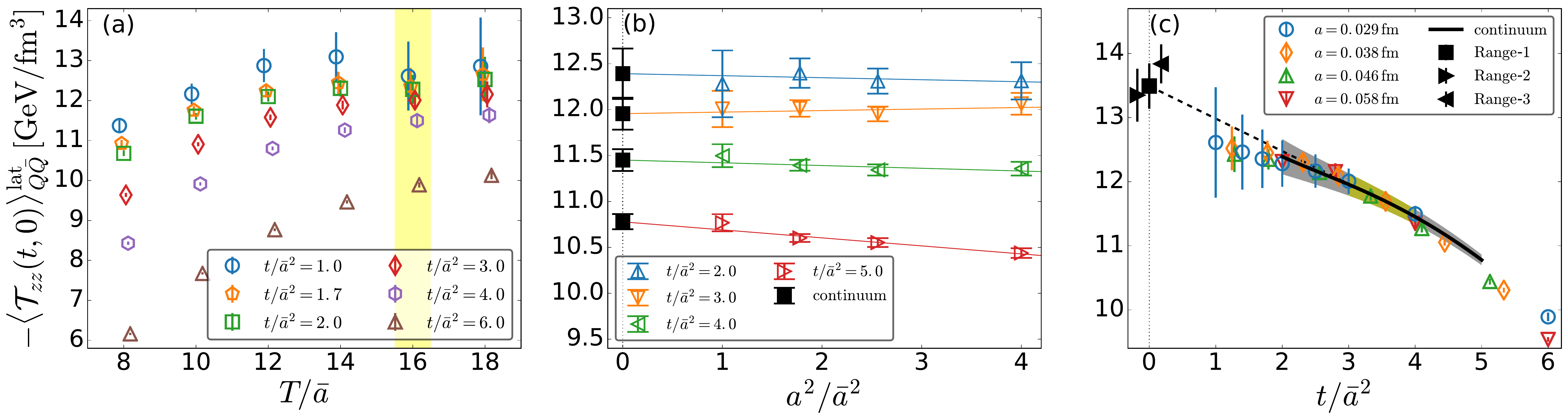}
 \caption{
 $zz$ component of the  stress tensor at $\vec{x}=0$,
 for various $a$ and $t$ with $R=0.46$~fm~\cite{Yanagihara:2018qqg}.
 (a) $\langle {\cal T}_{zz}(t,0)\rangle^{\rm lat}_{Q\bar{Q}}$
 as a function of $T/\bar{a}$ with $a= \bar{a}=0.029\ {\rm fm} $.
 To take the double limit, the data at
 $T/\bar{a}=16$ indicated by the yellow band is used.
 (b) Open symbols with errors denote
 $\langle {\cal T}_{zz}(t,0)\rangle^{\rm lat}_{Q\bar{Q}}$
 as a function of $a^2/\bar{a}^2$. The filled black symbols are
 the results of the $a\to 0$ limit for each $t$.
 (c) Open symbols with errors are
 $\langle {\cal T}_{zz}(t,0)\rangle^{\rm lat}_{Q\bar{Q}}$ as a function of
 $t/\bar{a}^2$ for different $a$. The solid line corresponds to the
 result of $a\to 0$ limit in the interval $2 \le t/\bar{a}^2 \le 5$ with shaded band being statistical
 error. The filled black symbols are the results of the $t\to 0$
 extrapolation from three different ranges of $t$.}
 \label{fig:mix}
\end{figure*}

We consider three $Q\bar{Q}$ distances ($R=0.46, 0.69, 0.92$ fm) 
shown in Table~\ref{table:param},
which are comparable to the typical scale of strong interaction.
While the largest $R$ is half the spatial lattice extent $aN_{\rm size}$
for the two finest lattice spacings, the finite size effects are 
known to be well suppressed even with this setting~\cite{Bali:1994de}.
A measure of the ground state saturation in  the $Q\bar{Q}$ system is 
characterized by
${P}(R, T) = {C_0 (R)  e^{-V(R)  T} }/{\langle W(R,T)\rangle_0}$ 
with the ground-state potential $V(R)$ and the ground-state overlap $C_0 (R)$
obtained at large  $T$ ~\cite{Bali:1994de}.
The data at $a=0.038$ fm shows $|1-{P} (R, T)| < {0.5} \%$ 
as long as $T > 0.19$ fm 
for  all $R$ in Table~\ref{table:param}. 
Employing $T\simeq0.46$~fm 
as shown in the last column of Table~\ref{table:param} ensures
the ground state saturation of the Wilson loop.
In order to avoid the discritization effect and
the over-smearing of the gradient flow~\cite{Kitazawa:2016dsl},
one has to
choose an  appropriate window of $t$ satisfying the 
condition  ${a}/{2} \lesssim  \rho  \lesssim L$, where
$\rho \equiv \sqrt{{2t}}$ is the flow radius and 
$L\equiv {\rm min} (|\vec{x}-\vec{R}_{+}|, |\vec{x}-\vec{R}_{-}|, T/2)$
is the minimal distance between $x_{\mu}=(\vec{x},0)$ and the Wilson loop.

\section{Stress Distribution on the Plane including Two Sources}

Here, we consider the stress distribution 
on the plane including two charges.
Shown in Fig.~\ref{fig:stress-distribution}~(a) is 
the two eigenvectors of the local stress tensor
around the two charges separated by $R=0.69\ {\rm fm}$
obtained on the finest lattice with $a=\bar{a}\equiv0.029\ {\rm fm}$ with fixed $t/{a}^2 = 2.0$.
The other eigenvector is perpendicular to the $y$-$z$ plane.
The eigenvector with negative (positive) eigenvalue 
is denoted by the red outward (blue inward) arrow
with its length proportional to $\sqrt{|\lambda_k |}$:
\begin{eqnarray}
 \red{\leftarrow} \hspace{-0.05cm} {\tiny{\circ}} \hspace{-0.05cm}  \red{\rightarrow} : \lambda_k<0, 
  \qquad  \blue{\rightarrow} \hspace{-0.05cm}  {\tiny{\circ}}  \hspace{-0.05cm}  \blue{\leftarrow}: \lambda_k>0.
\end{eqnarray}
Neighbouring volume elements are pulling (pushing) with each other
along the direction of red (blue) arrow.   
The spatial regions near $Q$ and $\bar{Q}$, which 
would suffer from over-smearing, 
are excluded in Fig.~\ref{fig:stress-distribution}. 
The direct analysis of the stress tensor
clearly reveals the spatial structure 
of the flux tube in a gauge invariant manner.
This is in contrast with the case of the classical electrodynamics 
with opposite charges shown in Fig.~\ref{fig:stress-distribution}~(b).

\section{Stress Distribution on the Mid-Plane between Two Sources}
Let us turn to the mid-plane between $Q$ and $\bar{Q}$.
As we take the double extrapolation in the following analysis,
we first show this procedure focusing on the mid point $\vec{x}=0$
as an example.
Fig.~\ref{fig:mix}~(a) shows the $T$ dependence of
$\langle {\cal T}_{zz}(t,0)\rangle_{Q\bar{Q}}^{\rm lat}$
for several values of $t$ 
with $a=0.029\ {\rm fm}\ (\equiv \bar{a})$.
The figure indicates that there exists a plateau
at large $T$, while it becomes narrower as $t$ becomes larger and
disappears at $t/\bar{a}^2 \sim 6$ 
(i.e., $\rho \sim \sqrt{12}\bar{a}=0.10$ fm).
The existence of the plateau means the ground state saturation, while
its violation comes from
the over-smearing effect.
From Fig.~\ref{fig:mix}, we thus conclude that
the over-smearing effect is not severe
with $T =16 \bar{a}=0.46$ fm at $t/\bar{a}^2 \leq 5$.
The same conclusion applies to all the other lattice spacings
in Table~\ref{table:param}.
On the other hand,
the lattice discretization effect is non-negligible due to
the under smearing
at $t/\bar{a}^2=2$ ($\rho=a=0.058$ fm) on the coarsest lattice.
Therefore, in the following analysis, we focus on the data in
the interval $2 \leq t/\bar{a}^2 \leq 5$
($2 \bar{a}  \leq \rho\leq \sqrt{10} \bar{a}$), which satisfies
${a}/{2} \lesssim  \rho  \lesssim L$ with margin.

In Fig.~\ref{fig:mix}~(b), we show the $a^2/\bar{a}^2$ dependence of 
$\langle {\cal T}_{zz}(t,0)\rangle_{Q\bar{Q}}^{\rm lat}$
for various $t$ by the open triangles. 
Here, we have taken the continuum limit with fixed $t$
which is shown by the filled black squares.

\begin{figure*}[t]
  \centering
  \includegraphics[width=0.95\textwidth,clip]{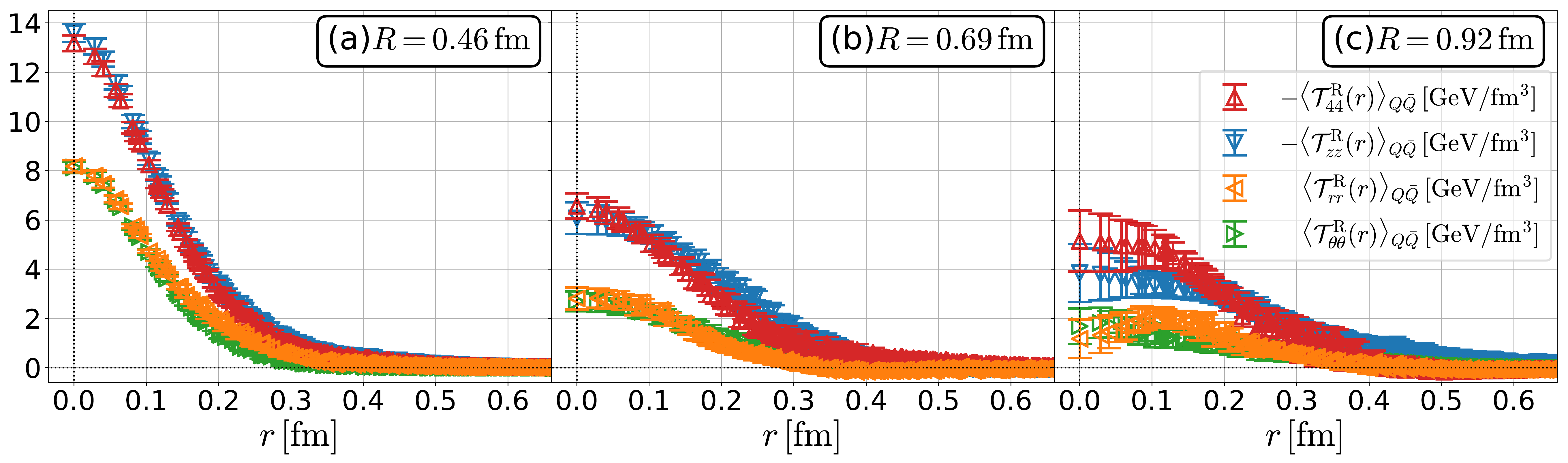}
  \caption{
EMT distribution on the mid-plane after the double limit
 $-\langle{\cal T}^{\rm R} _{cc}(r) \rangle_{Q\bar{Q}} $
 and $-\langle{\cal T}^{\rm R} _{44}(r) \rangle_{Q\bar{Q}} $
 in the cylindrical coordinate system for three different values of the $Q\bar{Q}$ distance $R$~\cite{Yanagihara:2018qqg}.
  \label{fig:mid}
  }
\end{figure*}

In Fig.~\ref{fig:mix}~(c), 
we show the values of $\langle {\cal T}_{zz}(t,0)\rangle_{Q\bar{Q}}^{\rm lat}$
for various values of  $a$ and $t/\bar{a}^2$ by the open symbols,
and the result of the continuum extrapolation in the interval
$2 \le t/\bar{a}^2 \le 5$
by the black solid line with the shaded error band.
The $t\rightarrow 0$ limit is carried out
by using the values in the continuum limit.
We have carried out the $t\rightarrow 0$ extrapolation
with three fitting windows in order to estimate the systematic errors: 
$3 \leq t/\bar{a}^2 \leq 4$ (Range-1), 
$2 \leq t/\bar{a}^2 \leq 4$ (Range-2) and 
$3 \leq t/\bar{a}^2 \leq 5$ (Range-3).
The final results of $\langle {\cal T}_{zz}^{\rm R}(0)\rangle_{Q\bar{Q}}$
after the double limit
are shown by the filled black symbols. The dashed line
represents the extrapolation with Range-1.

\begin{figure}
 \centering
 \includegraphics[width=0.47\textwidth,clip]{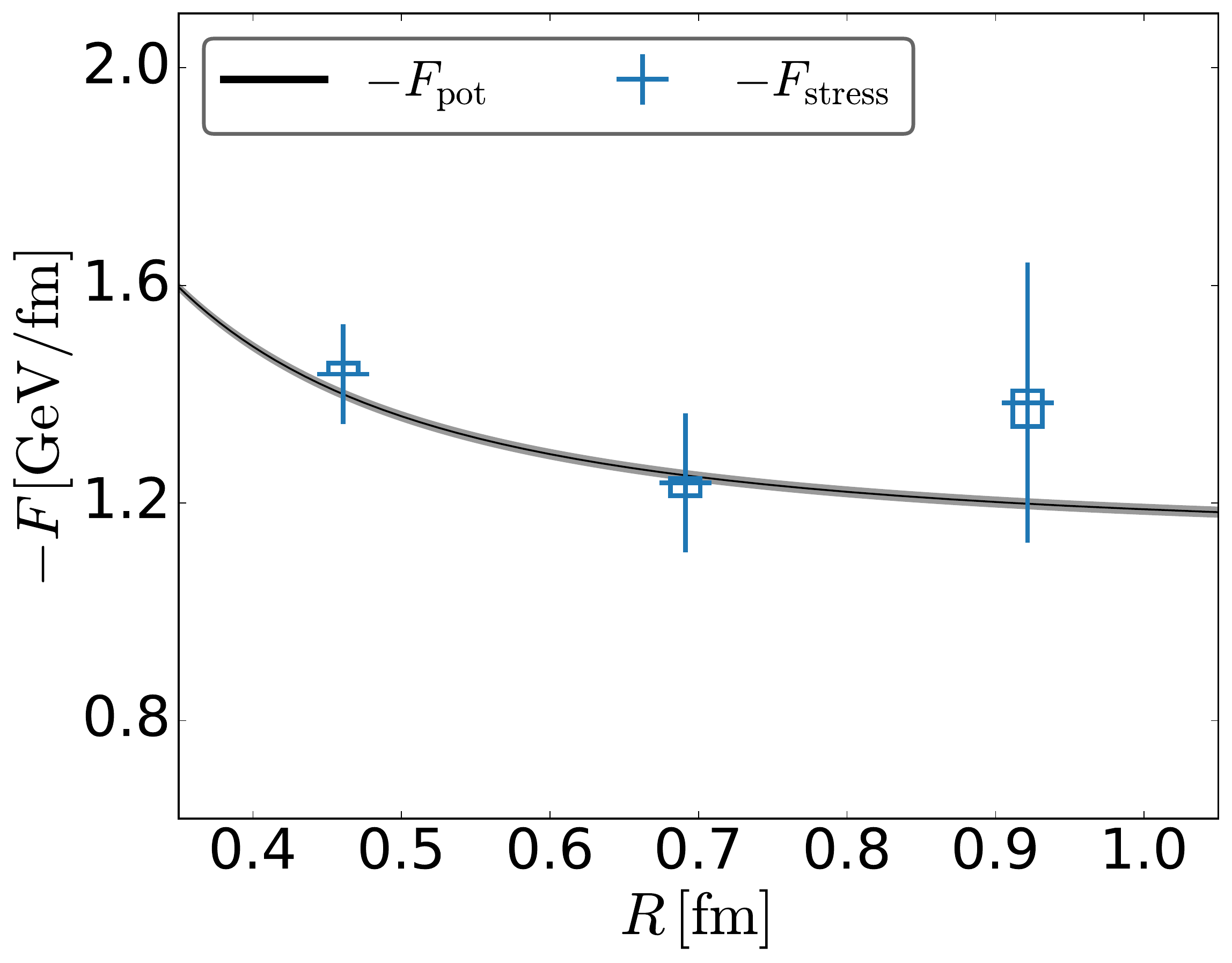}
 \caption{
 $R$ dependence of the $Q\bar{Q}$ forces,
 $-F_{\rm stress}$ and $-F_{\rm pot}$~\cite{Yanagihara:2018qqg}.
 Error bars and rectangular boxes for the latter represent
 the statistical and systematic errors,
 respectively.
 \label{fig:R}
 }
\end{figure}

Let us now turn to the whole mid-plane.
We use the cylindrical coordinate system $c=(r,\theta, z)$ with
$r=\sqrt{x^2+y^2}$ and $0 \le \theta < 2 \pi$. 
On the mid-plane, because of the cylindrical symmetry and
the parity symmetry with regard to $z$ axis, 
EMT is diagonalized as 
$\langle {\cal T}_{c c'}(t,x) \rangle_{Q\bar{Q}}^{\rm lat} =
{\rm diag} ( \langle{\cal T}_{rr} (t,r)\rangle_{Q\bar{Q}}^{\rm lat},
\langle {\cal T}_{\theta\theta} (t,r)\rangle_{Q\bar{Q}}^{\rm lat},
\langle {\cal T}_{zz} (t,r)\rangle_{Q\bar{Q}}^{\rm lat})$.
To take the continuum limit,
we need the data at the same $r$ for different $a$.
We consider the values of $r$ at which the lattice data are 
available on the finest lattice. To obtain the data 
at these $r$ on lattices with other $a$, we interpolate the 
lattice data $\langle {\cal T}_{c c}(t,r) \rangle_{Q\bar{Q}}^{\rm lat}$
 and $\langle {\cal T}_{44}(t,r) \rangle_{Q\bar{Q}}^{\rm lat}$
with the commonly used functions
to parametrize the transverse profile of the flux tube:
$f_{\rm Bessel} (r)= A_0 K_0\big(\sqrt{Br^2+C}\big)$
with the 0th-order modified Bessel function $K_0(x)$~\cite{Clem:1975}
and $f_{\rm exp} (r)= (A_0+A_1r^2) e^{(-2\sqrt{r^2+B^2}+2B)/C}$~\cite{Cardoso:2013lla}.
The $t \to 0$ limit is taken in the same way as explained above.

In Fig.~\ref{fig:mid}, we show the $r$ dependence of the resulting EMT.
From the figure, one finds several noticeable features:
(i) Approximate degeneracy between temporal
and longitudinal components 
is found for a wide range of $r$:  
$ \langle {\cal T}^{\rm R}_{44}(r)  \rangle_{Q\bar{Q}}  \simeq
\langle {\cal T}^{\rm R}_{zz}(r) \rangle_{Q\bar{Q}} < 0 $.
We also find 
$ \langle {\cal T}^{\rm R}_{rr}(r)  \rangle_{Q\bar{Q}}  \simeq
\langle{\cal T}^{\rm R}_{\theta \theta}(r)  \rangle_{Q\bar{Q}} >0  $, which 
does not have an obvious reason except at $r=0$.
(ii) The scale symmetry broken in the YM vacuum (the trace anomaly) is
partially restored inside the flux tube,
which is related to the non-zero trace anomaly,  
$ \langle {\cal T}^{\rm R}_{\mu \mu}(r) \rangle_{Q\bar{Q}} =
\langle {\cal T}^{\rm R}_{44}(r) +  {\cal T}^{\rm R}_{zz}(r) +
{\cal T}^{\rm R}_{rr}(r)  +  {\cal T}^{\rm R}_{\theta \theta}(r) \rangle_{Q\bar{Q}} <  0 $.
This is in sharp  contrast to  the case of classical electrodynamics;
${\cal T}_{44}(r)={\cal T}_{zz}(r)=-{\cal T}_{rr}(r)=
-{\cal T}_{\theta \theta}(r)$ and    ${\cal T}_{\mu \mu}(r)=0$ for all $r$.
(iii) Each component of EMT at $r=0$ becomes smaller as
$R$ becomes larger, while the transverse radius of the flux tube,
typically about $0.2$~fm, seems to be wide
for large $R$~\cite{Cardoso:2013lla,Luscher:1981}, 
although the statistics is not enough to discuss the radius quantitatively.

Finally, we show a non-trivial consistency check of our analysis.
We can define the force acting on the charge located at $z>0$
in two different ways, which should give the same result.
First, the force is defined from the $Q\bar{Q}$ potential
as $F_{\rm pot}=-dV(R)/dR$. Second, one can 
also obtain the force 
by the surface integral of the stress-tensor 
surrounding the charge, 
$F_{\rm stress} = - \int  \langle {\cal T}_{zj}(x) \rangle_{Q\bar{Q}}\  dS_j$.
For $F_{\rm pot}$,
we fit the numerical data of  $V(R)$ calculated from the Wilson loop
at $a=0.038$ fm with the Cornell parametrization.
Note that this lattice spacing is 
already close to the continuum limit~\cite{Bali:2000gf}.
For $F_{\rm stress}$, the surface integral is performed on the mid-plane:
$F_{\rm stress} = 2 \pi \int_0^{\infty} \langle {\cal T}_{zz}(r)
\rangle_{Q\bar{Q}}  \ r dr $.
Here $\langle {\cal T}_{zz}(r) \rangle_{Q\bar{Q}}$ is obtained
by fitting the data in Fig.~\ref{fig:mid}
with either $f_{\rm Bessel}(r)$ or  $f_{\rm exp}(r)$.
In Fig.~\ref{fig:R}, $- F_{\rm pot}$ and $- F_{\rm stress}$
thus obtained are shown by the solid line and the
horizontal bars, respectively.
For $-F_{\rm stress}$, we take into account not only the statistical error but
also the systematic errors from the double limit and the choice of
fitting function,
$f_{\rm Bessel,exp}(r)$.
The agreement between the two quantities within the errors
is a first numerical evidence that the ``action-at-a-distance''
$Q\bar{Q}$ force can be described
by the  local properties of the stress tensor in YM theory.

\section{Summary and Outlook}
We have explored the spatial distribution of EMT
around the $Q\bar{Q}$ system in SU(3) lattice gauge theory. 
The YM gradient flow plays a crucial role to define EMT on the lattice.
We have investigated
the stress-tensor distribution on the mid plane by taking
the double limit.
The linear confining behavior of the $Q\bar{Q}$ potential at long distances
is obtained
by the surface integral of the stress tensor.

There are interesting applications 
of this study, such as 
the generalization to full QCD with the QCD flow
equation~\cite{Makino:2014taa}
and the analyses of the $QQQ$ system and the $Q\bar{Q}$ system
at finite temperature.

\section*{Acknowledgement}
The numerical simulation was carried out on IBM System Blue Gene Solution
at KEK under its Large-Scale Simulation Program (No.~16/17-07).
This work was supported by JSPS Grant-in-Aid for Scientific Researches, 
17K05442, 25287066 and 18H05236.

\end{document}